# Multi-state electromagnetic phase modulations in NiCo$_2$O$_4$ through cation disorder and hydrogenation

*Xuanchi Zhou [1, 2] †\*, Xiaohui Yao [1] †, Shuang Li [1], Xiaomei Qiao [1], Jiahui Ji [1],*

*Guowei Zhou [1, 2]\*, Huihui Ji [1, 2], Xiaohong Xu [1, 2] \**

[1] *Key Laboratory of Magnetic Molecules and Magnetic Information Materials of Ministry of Education & School of Chemistry and Materials Science, Shanxi Normal University, Taiyuan, 030031, China*

[2] *Research Institute of Materials Science, Shanxi Key Laboratory of Advanced Magnetic Materials and Devices, Shanxi Normal University, Taiyuan 030031, China*

\*Authors to whom correspondence should be addressed: *xuanchizhou@sxnu.edu.cn (X. Zhou)*, *zhougw@sxnu.edu.cn (G. Zhou)*, and *xuxh@sxnu.edu.cn (X. Xu)*.

*† X. Zhou, and X. Yao contributed equally to this work.*

**Abstract**

One focal challenge in engineering low-power and scalable all-oxide spintronic devices lies in exploring ferromagnetic oxide material with perpendicular magnetic anisotropy (PMA) and electronic conductivity while exhibiting tunable spin states. Targeting this need, spinel nickel cobaltite ($NiCo_2O_4$, NCO), featured by room-temperature ferrimagnetically metallic ground state with strong PMA, emerges as a promising candidate in the field of oxide spintronics. The cation distribution disorder inherent to NCO renders competing electromagnetic states and abnormal sign reversal of anomalous Hall effect (AHE), introducing an additional freedom to adjust electromagnetic transports. Here, we unveil multi-state electromagnetic phase modulations in NCO system through controllable cation disorder and proton evolution, extensively expanding electromagnetic phase diagram. The cation disorder in NCO tunable by growth temperature is identified as a critical control parameter for kinetically adjusting the proton evolution, giving rise to intermediate hydrogenated states with chemical stability. Hydrogen incorporation reversibly drives structural transformation and electromagnetic state evolutions in NCO, with rich spin-dependent correlated physics uncovered by combining the AHE scaling relation and synchrotron-based spectroscopy. Our work not only establishes NCO as a versatile platform for discovering spin-dependent physical functionality but also extends the horizons in materials design for state-of-the-art spintronic devices harnessing magneto-ionic control and inherent cation disorder.

## 1. Introduction

The pursuit of low-power and scalable spintronic devices harnessing both the charge and spin degrees of freedom of electrons hinges critically on seeking for the magnetic material layer with strong perpendicular magnetic anisotropy (PMA), for example, CoFeB, FePt and [Co/Pt]$_n$.[1-5] Beyond that, oxide spintronics emerges as a promising paradigm for exploring exotic spin-dependent physical functionality and phenomena, owing to their exceptionally tunable spin states and complicated interplay among multiple degrees of freedom.[6-8] Nevertheless, the promise of oxide spintronics is fundamentally restricted by the scarcity of intrinsic oxide counterparts to traditional ferromagnets that integrates room-temperature ferromagnetic order, PMA, and metallicity.[9] Within this landscape, ferrimagnetic spinel nickel cobaltite (NiCo$_2$O$_4$, NCO), featured by the robust PMA and metallic conductivity, is identified as a compelling candidate for engineering all-oxide spintronic devices.[10-11] The inverse spinel structure of NCO, characterized by Co exclusively occupying the tetrahedral ($T_d$) sites and Ni/Co equally sharing the octahedral ($O_h$) sites, inherently hosts the cation disorder and mixed valency. The ferrimagnetic ground state of NCO with the Curie temperature ($T_c$) of ~390 K results from the antiparallel magnetic moments of Ni and Co, with the metallic conductivity primarily governed by the high-spin Ni$^{3+}$ state in the $O_h$-sublattice.[12-13] In addition, the compatible growth temperature of NCO material (e.g., ~300 ºC) with traditional semiconductor processing,[14] along with tunable spin states via external stimuli, has captured considerable attention in the spintronics community.

One focal challenge persists in the reversible control over the electromagnetic states in NCO system in a more robust energy landscape, a cornerstone of spin valve and magnetic tunnel junction in modern spintronics. In recent years, the magneto-ionic control over the ion-charge-spin interactions in transition metal oxide system offers a rewarded pathway for modulating the electromagnetic states in a more reversible and controllable fashion.[15-20] The incorporation of mobile hydrogens acts as an electron donor, injecting carriers into the conduction band to drive electronic phase modulations, while simultaneously adjusting the spin configuration and magnetic exchange interactions.[21-23] Hydrogen-related electromagnetic phase transformations are previously demonstrated in NCO system, transitioning ferrimagnetically metallic ground state to antiferromagnetic insulator.[24-25] The incorporation of mobile hydrogens can be kinetically adjusted by the local disorder and microstructure design of oxide lattice, enabling the rational design in exotic quantum states.[26] The inherent cation distribution disorder in NCO system, controllable by deposition temperature and film thickness, is poised to introduce an additional freedom to manipulate electromagnetic states through proton evolution.[27] Consequently, exotic electromagnetic states are expected to emerge within the hydrogen-related phase diagram of NCO, advancing protonic device applications in oxide spintronics.

In this work, we demonstrated hydrogen-related electromagnetic phase modulations in NCO system that fosters diverse electromagnetic states with robust chemical stability and reversibility, the kinetics of which is facilely tunable by the cation distribution disorder. Hydrogen doping drives multi-state phase modulations from a ferrimagnetic metal to an antiferromagnetic insulator in NCO, featuring intermediate hydrogenated states, with the underlying mechanism unveiled by synchrotron spectroscopy. The scaling relation between the anomalous Hall conductivity and longitudinal conductivity unveils a reduced contribution from intrinsic Berry curvature in NCO through hydrogenation. Our findings not only discover abundant hydrogenated states in NCO system, with great potential for spintronic devices leveraging proton evolution, but also clarify how the incorporated hydrogens affect spin-dependent transport behaviors.

2. **Results and discussion**

Motivated by adjustable cation distribution disorder in NCO system by deposition temperature, high-quality epitaxial NCO thin films were grown on single-crystalline MgAl$_2$O$_4$ (MAO) (001) substrate through the laser molecular beam epitaxy (LMBE) technique under different deposition temperatures ranging from 325 °C to 425 °C. The intrinsic cation distribution disorder of inverse spinel NCO is schematically illustrated in Figure 1a, described by $[Co_x^{2+}Co_{1-x}^{3+}]_{T_d}[Co^{3+}Ni_x^{2+}Ni_{1-x}^{3+}]_{O_h}O_4^{2-}$, where Co cations exclusively occupy the $T_d$ sites, with the equal filling of Ni and Co cations in the $O_h$ sites.[28] Therefore, such the disorders of cation distribution and mixed valence states give rise to competing electronic and magnetic states in the phase diagram of NCO, while adjusting the proton evolution kinetics. Given an enlarged *a*-axis lattice constant of NCO bulk ($a_{pc, NCO}$=8.114 Å) in comparison with the MAO substrate ($a_{pc, MAO}$=8.083 Å) with a small lattice mismatching of -0.38 %, the NCO film is expected to epitaxially grown on the *c*-plane MAO template, with an *in-plane* compressive distortion. This understanding is further confirmed by respective reciprocal space mapping (RSM) around the MAO (226) reflection (Figure 1b), where the identical *in-plane* vector (e.g., $Q_\parallel$) between the NCO film grown at 325 °C and MAO substrate collaborates the epitaxial growth of NCO/MAO (001) heterostructure. Furthermore, as-observed enlarged *cross-plane* vector (e.g., $Q_\perp$) for the MAO substrate with respect to the NCO film confirms the *in-plane* biaxial compressive distortion within the grown NCO film, resulting from the *a*-axis lattice mismatch of approximately -0.38 %.

The structural evolution in NCO film via cation distribution disorder is verified by comparing their X-ray diffraction (XRD) patterns of the grown NCO films under diverse deposition temperatures (Figure 1c). The diffraction peak associated with the (004) plane of NCO films just appears adjacent to the (004) diffraction peak of MAO substrates, identifying an *out-of-plane* preferential orientation induced by epitaxial template. Enlarging the deposition temperature of NCO films is prone to induce a more disordered distribution of nickel and cobalt cations within the lattice of inverse spinel NCO. A systematic shift of the (004) diffraction peak of NCO films toward lower angles is observed with increasing the growth temperature, indicating the lattice expansion along the *out-of-plane* direction (Figure S1). Such the structural expansion is likely attributed to enhanced cation disorder within the NCO lattice that drives the emergence of oxygen vacancies at higher deposition temperature, inducing the lattice expansion. Atomic force microscopy (AFM) measurements determine the NCO film thickness to be approximately 22 nm and reveal a smooth surface with a root-mean-square roughness below 0.55 nm, which can be further reduced by increasing the growth temperature (Figure S2).

Tunable cation distribution disorder in NCO system provides a fertile ground for modulating the resulting electromagnetic transport behaviors, unlocking competing electronic and magnetic states in the phase diagram. It is found that the NCO/MAO heterostructure grown at 325 °C showcases a metallic transport behavior with a positive magnitude of $d$R/$d$T from room temperature down to ~50 K, below which a resistivity upturn is observed, primarily attributed to the enhanced electron-electron interactions in low-temperature region (Figures 1d and S3).[29] Nevertheless, the electronic phase transition from metal to semiconductor is achieved for NCO film via elevating the growth temperature to 425 °C, evidenced by the respective temperature dependences of material resistivity ($\rho$-$T$). The elevation in the material resistivity of NCO film with the deposition temperature, as well as the observed electronic phase transition, aligns well with aforementioned cation-disorder-driven structural evolution. The suppression in high-spin $Ni^{3+}$ state of the grown NCO film via the formation of oxygen vacancies under higher deposition temperature deteriorates the electron-itinerant ground state.

The intimate coupling between lattice, charge and spin degrees of freedom in NCO system not only underpins the rich correlated physics, but also enables the synergistic control over the magnetic ordering via cation disorder. The grown NCO film at 325 °C exhibits a robust PMA, as evidenced by comparing the magnetization hysteresis (*M-H*) loops measured at 300 K along the *in-plane* and *out-of-plane* directions (Figure 1e). Furthermore, temperature-dependent magnetization (*M-T*) curve unveils the magnitude of $T_c$ for the NCO film grown at 325 °C exceeding 380 K (Figure S4), aligning well with the reported bulk value. The square-shaped *M-H* loops measured with the magnetic field applied normal to the film plane demonstrate the strong PMA in the NCO films grown at 325 °C and 375 °C, which, however, degrades significantly with increasing the deposition temperature to 425 °C (Figures 1f and S5). In all, a higher growth temperature drives more disordered cation distribution in NCO film with oxygen deficiency, dramatically degrading the ferrimagnetically metallic ground state with PMA. The above competing electromagnetic states in NCO system establish growth-temperature-driven cation disorder as a powerful tuning knob for tailoring the structural evolution and electromagnetic transport behaviors.

To realize the reversible control over the electromagnetic states in NCO system, hydrogenation was employed via using the hydrogen spillover strategy. Utilizing the sputtered Pt catalyst, the energy barrier for dissociating the $H_2$ molecule into protons and electrons at the triple phase boundary is significantly reduced, promoting the proton evolution in NCO (Figure S6).[30] Here, the NCO film grown at 325 °C, featured by well-defined ferrimagnetically metallic ground state with PMA, is identified as a model system to investigate hydrogen-related electromagnetic phase modulations. Hydrogenation of NCO at 70 °C for 10 min leads to the emergence of a new diffraction peak related to the hydrogenated phase, concurrent with the preservation of the original NCO signature (Figure 2a). Further elevating the hydrogenation temperature and extending the hydrogenation period (e.g., 100 °C, 1 h) transforms the intermediate state into a fully hydrogenated state, annihilating the pristine NCO diffraction peak. The incorporation of interstitial hydrogens induces the lattice expansion of NCO film along the *out-of-plane* direction, attributed to the O-H interactions. The anisotropy in hydrogen-induced structural phase transformation of NCO films is evident by respective RSM spectra (Figure S6). The *in-plane* lattice of NCO film is tightly locked by the epitaxial MAO template, contrasting sharply with the lattice expansion along the *out-of-plane* direction, where the magnitude of $Q_\perp$ is reduced via proton evolution. In particular, the lattice framework of NCO parent remains intact through hydrogenation, a hallmark of topotactic phase transformation, which is consistent with similar Raman spectra (Figure S7).

Apart from the structural phase transformation, hydrogen incorporation drives electromagnetic phase modulations in NCO system. The incorporation of mobile hydrogens depresses the metallic conductivity of NCO, accompanied by the progressive elevation in the material resistivity (Figure 2b). Meanwhile, the PMA of pristine NCO is suppressed through hydrogenation (Figure 2c), with the reduction in the saturation magnetization ($M_s$) (Figure S8). It is worthy to note that hydrogen-driven electromagnetic phase modulations are amplified through an intense hydrogenation at 100 °C for 1 h, in agreement with their XRD results. To uncover the changes in chemical environment of NCO through hydrogenation, soft X-ray absorption spectroscopy (sXAS) was conducted, as the results shown in Figures 2d-2f, respectively. It is widely reported that pristine NCO features a mixed valency, the disorder of which is characterized by the mixed $Ni^{3+}/Ni^{2+}$ in $O_h$ sites and $Co^{3+}/Co^{2+}$ states in both $O_h$ and $T_d$ sites. The Ni *L*-edge and Co *L*-edge arising from the $2p\rightarrow3d$ transitions of transition metals serve as an effective indicator for reflecting the variations in the valence states of nickel and cobalt, providing spectroscopic evidence for clarifying the physical picture underneath hydrogen-related phase modulations. Performing the hydrogenation leads to the reduction in the relative intensity of peak A associated with high-valence $Ni^{3+}$ state (Figure 2d).[31] It is well-known that the high-spin $Ni^{3+}$ state ($3d^7$) in NCO facilitates the carrier delocalization near the Fermi level, accounting for the metallic conductivity of NCO.[32] The introduction of hydrogens meanwhile donates electron carriers to reduce the valence state of nickel from $Ni^{3+}$ to $Ni^{2+}$ and depress the $Ni^{3+}$-$O^{2-}$-$Ni^{2+}$ double exchange interaction, evidenced by the sXAS spectra of Ni *L*-edge, leading to the hydrogen-induced carrier localization. In addition, the reduced valence state of cobalt in hydrogenated NCO is further evidenced by the emergence of peaks B and B' in the Co *L*-edge spectra, characteristic of low-valence $Co^{2+}$ state (Figure 2e). Therefore, the antiferromagnetic coupling between Co ions occupying $O_h$ and $T_d$ sites via superexchange interactions is directly responsible for suppressing the net magnetic moment, thereby diminishing the ferrimagnetic ordering, according to previous reports.[24, 33] The PMA of NCO arising from the spin-lattice coupling and the broken cubic symmetry is correlated with the $d_{x^2-y^2}$ states in $T_d$-site Co ions and orbital angular momentum along the $z$ axis.[34] Consequently, hydrogenation is also expected to reduce the orbital angular momentum along the $z$ axis and the coupling strength between $|d_{x^2-y^2}, S_z = 1/2\rangle$ state and $|d_{xy}, S_z = 1/2\rangle$ state through hydrogen-associated band-filling control. Meanwhile, the hydrogenation results in the depression of the orbital hybridization between Co-3$d$/Ni-3$d$ orbital and O-2$p$ orbital, underpinning the hydrogen-induced carrier localization (Figure 2f).

To clarify the chemical stability and reversibility, the temporal evolutions in structural transformation, electrical transports and valence state changes of hydrogenated NCO at 100 °C for 1 h via air exposure are shown in Figure 3. The characteristic diffraction peak of the hydrogenated state remains chemically stable for 39 days in air but reversibly recovers to the pristine state after 52 days (Figure 3a), the stability of which surpasses the intermediate counterpart (e.g., hydrogenated at 70 °C for 10 min) (Figure S9). Nevertheless, the residual hydrogens in the deeper layer of NCO lattice impedes the complete recovery to the initial state, as collaborated by their $\rho$-$T$ tendencies (Figure 3b). Despite a reversible tendency is confirmed after air exposure, the resistivity of the NCO film following 52 days in air remains lower than that of its pristine state. In addition, the shoulder peak associated with high-valence $Ni^{3+}$ state in the Ni $L_3$-edge sXAS spectra is revived through ambient exposure for 52 days, confirming the elevation in the nickel valence state (Figure 3c). The recovery of the characteristic anomalous Hall effect (AHE) provides definitive evidences for the reversibility in the magneto-ionic control over electromagnetic states in NCO system (Figure 3d). After exposing to the air for 52 days, the Hall resistance ($R_{xy}$) of hydrogenated NCO shows a robust switching hysteresis under an *out-of-plane* magnetic field, signaling the recovery of typical AHE. This observation starkly differs from the hydrogenated NCO counterpart, in which situation a diminished AHE behavior verifies the extensively depressed ferrimagnetic ordering via proton evolution. The ultrahigh mobility of incorporated hydrogens, together with the chemical potential to extract the hydrogens from NCO lattice via the Pt catalyst, engenders a spontaneous dehydrogenation process in NCO system when exposing to the air for 52 days (Figure 3e). However, the hydrogenated states in the NCO system remain stable under ambient exposure for 39 days. Such the relatively stable but reversible hydrogenated states in NCO system benefit the protonic device applications in oxide spintronics.

The AHE as a pivotal electronic transport behavior, driven by band Berry phase or impurity scattering, provides a powerful tool for probing the magneto-transport properties in magnetic materials.[35-37] For pristine NCO film grown at 325 °C, a square-shaped hysteresis with a negative coefficient is clearly identified, providing compelling evidences for intrinsic ferrimagnetism (Figure 4a). A mild hydrogenation progressively suppresses the AHE signal in NCO, as further exemplified by temperature-dependent AHE measurements (Figure 4b). Nevertheless, excessive hydrogenation at 100 °C for 1 h leads to the complete quenching of AHE. Moreover, beyond the hydrogen-triggered reduction of $R_{xy}$, the coercive field ($H_c$) of hydrogenated NCO extracted from the AHE loops showcases an exponential growth with the decreasing temperature, starkly differing from the pristine counterpart that exhibits a similar $H_c$ (Figure 4c). This distinct temperature tendency of $H_c$ in hydrogenated NCO likely stems from domain-wall pinning caused by the enhanced disorder introduced via hydrogenation.[10] To gain deeper insight into how incorporated hydrogens modify the ferromagnetic response by spin-polarized charge carriers, the scaling relation between the anomalous Hall conductivity ($\sigma_{xy}$) and the longitudinal conductivity ($\sigma_{xx}$) is further analyzed (Figures 4d and S10). Pristine NCO resides on the moderately dirty regime, the power-law relation of which can be described by $\sigma_{xy} = A\sigma_{xx}^{1.6}+B$.[38-39] Therefore, the AHE is determined by both intrinsic band intrinsic contribution (coefficient B) and scattering-dependent extrinsic contribution arising from the skew scattering and side jump ($A\sigma_{xx}^{1.6}$). On the basis of fitting the scaling relation, the absolute magnitude of B associated with the intrinsic Berry curvature is significantly reduced through hydrogenation, pointing to a heightened role of scattering-dependent extrinsic mechanism (Figure 4e). This shift toward a disorder-dominated scattering regime mirrors the effect of increasing cation disorder in pristine NCO by raising the deposition temperature (Figures S11-S12). Hydrogen incorporation not only adjusts the electromagnetic transport behaviors of NCO, but also actively reconfigures the underlying spin-dependent physical mechanisms.

The electromagnetic transport behaviors in NCO system are tunable by using the cation distribution disorder, resulting in competing electromagnetic states, as exemplified by the thickness- or temperature-dependent AHE sign reversal.[27, 40] Alternatively, cation distribution disorder is expected to introduce a new freedom to manipulate the hydrogen-related phase transitions in NCO via adjusting associated migration kinetics. This understanding is further confirmed by comparing the hydrogen-induced structural transformation and electromagnetic state evolutions within NCO system under hydrogenated at 100 ºC for 1 h (Figures 4f-4h). The cation disorder in NCO lattice induced by higher deposition temperatures acts as a kinetic barrier to hydrogen incorporation, generally suppressing hydrogen-driven structural and electromagnetic evolutions. Furthermore, for NCO grown at 375 ºC, hydrogenation drives not merely an increase in material resistivity but a full metal-to-semiconductor phase transition, contrasting sharply with the aforementioned counterpart grown at 325 ºC (Figure S13). In contrast to the evident magneto-ionic control of ferrimagnetism and PMA in NCO deposited at 375 ºC, NCO films grown at the higher temperature of 425 ºC show no obvious regulation upon hydrogenation, as their electromagnetic ground states are already perturbed via temperature-driven cation disorder (Figures S14-S15). Therefore, hydrogen-driven phase transformations are blocked by the 'saturated' cation disorder inherent to NCO films deposited at 425 ºC. Analogous chemical stability and reversibility are also observed for NCO films hydrogenated at either 375 ºC or 425 ºC (Figure S16).

Hydrogen doping offers a new paradigm to explore new spin states and exotic physical functionality in NCO system, transcending traditional paradigm. Hydrogen-related electron-doping Mottronics reduces the valence states of nickel and cobalt in NCO, depressing the ferrimagnetic ordering with PMA and metallic conductivity. Meanwhile, incorporated hydrogens act as scattering centers that suppress the intrinsic contribution of Berry curvature to the anomalous Hall conductivity, underpinning rich spin-dependent correlated physics. Benefiting from the delicate design of hydrogenation conditions, diverse hydrogenated states emerge in the hydrogen-related electromagnetic phase diagram, with high chemical stability and reversibility. In NCO films grown at 325 ºC, a mild hydrogenation (e.g., 70 ºC for 10 min) produces an intermediate state where electron itinerancy and the AHE are preserved, despite a reduction in PMA. After sequential hydrogenation at 100 ºC for 1 h, NCO loses its AHE and PMA but remains metallic, demonstrating an intriguing decoupling of ferrimagnetic ordering from electronic conductivity. The cation distribution disorder in NCO system, tunable by deposition temperature, introduces an additional freedom to adjust the electromagnetic ground states and proton evolution. An enhanced cation disorder in NCO lattice deposited at a higher temperature readily perturbs the ferrimagnetically metallic ground state with robust PMA, while in turn kinetically impeding subsequent hydrogen-related electromagnetic phase modulations. Our findings establish an emerging paradigm for accessing new electromagnetic states, with great potential for low-power spintronics, while deepening the understanding of spin-dependent transport behaviors.

## 3. Conclusion

In summary, we identified cation disorder as a powerful tuning knob for adjusting electromagnetic phase modulations in NCO system, delicately delivering diverse electromagnetic states. The cation disorder as a critical control parameter readily attenuates the ferrimagnetically metallic ground state of NCO, while kinetically hindering the proton evolution kinetics. Hydrogen-related filling-controlled Mottronics suppresses the electronic conductivity of NCO governed by high-spin $Ni^{3+}$ state and ferrimagnetic ordering, giving rise to intermediate states with robust chemical stability and reversibility. The incorporation of interstitial hydrogens reduces intrinsic contribution from Berry curvature to the anomalous Hall conductivity in NCO, deepening the understanding of how hydrogen doping affects spin-dependent physical pictures. The NCO system under hydrogenation hosts a rich spectrum of physical phenomena, exemplified by the decoupling between electron itinerancy and ferrimagnetic ordering. The present work not only extensively expands hydrogen-related electromagnetic phase diagram, with potential for advancing low-power spintronics, but also extends the horizons in material designs for unlocking exotic spin-related physical functionality and quantum states.

## 4. Experimental Section

*Fabrication of the grown NCO films:* Epitaxial NiCo$_2$O$_4$ (NCO) thin films were grown on single-crystalline (001)-oriented MgAl$_2$O$_4$ (MAO) substrates using laser molecular beam epitaxy (LMBE). The film deposition was performed at substrate temperatures ($T_{\text{sub.}}$) ranging from 325 °C to 425 °C. The oxygen partial pressure, target-to-substrate distance, and laser fluence were optimized as 20 Pa, 45 mm, and 1.0 J cm$^{-2}$, respectively. After the film deposition, all the grown NCO films were subjected to an *in-situ* post-annealing treatment in an oxygen-rich atmosphere at an oxygen partial pressure of 1.5×10$^4$ Pa for 30 minutes to minimize oxygen vacancies and improve the crystallinity. The NCO/MAO (001) heterostructures were subsequently cooled to room temperature under the identical oxygen partial pressure. Before the hydrogenation, the 20 nm-thick platinum dots were sputtered into the surface of the grown NCO films via exploiting the magnetron sputtering technique. Finally, based on the hydrogen spillover strategy, as-made Pt/NCO/MAO (001) heterostructures were annealed in a 5 % H$_2$/Ar forming gas for effectively realizing the hydrogenation.

*Material characterizations:* The crystal structures of the grown NCO films were probed via using the X-ray diffraction (XRD) (Rigaku, Ultima IV) and Raman spectra (HORIBA, HR Evolution). The epitaxial growth of the grown NCO film is identified by using the reciprocal space mapping (RSM) (Rigaku, Ultima IV). The surface topology and film thickness of the grown NCO films is identified by using the atomic force microscope (AFM) (Bruker, Dimension Icon). The electronic structure of NCO films was further explored through soft X-ray absorption spectroscopy (sXAS) analysis, as conducted at the Shanghai Synchrotron Radiation Facility (SSRF) on beamline BL08U1A. Magnetic properties were measured using a superconducting quantum interference device (SQUID) (Quantum Design). Temperature dependent resistivity of the deposited NCO films was measured by using a commercial physical property measurement system (PPMS) (Quantum design). The anomalous Hall effect (AHE) of the grown NCO films was investigated by using the commercial Keithley 6221 and 2182 system.

**Declaration of competing interest**

The authors declare no conflict of interest.

**Acknowledgements**

This work was supported by the National Key R&D Program of China (No. 2025YFA1411100), the National Natural Science Foundation of China (Nos. 52401240, 52471203, U24A6002, and 12174237), Fundamental Research Program of Shanxi Province (No. 202403021212123), Scientific and Technological Innovation Programs of Higher Education Institutions in Shanxi (No. 2024L145), and Shanxi Province Science and Technology Cooperation and Exchange Special Project (No. 202404041101030). The authors also acknowledge the beam line BL08U1A at the Shanghai Synchrotron Radiation Facility (SSRF) (https://cstr.cn/31124.02.SSRF.BL08U1A) and the beam line BL12B-b at the National Synchrotron Radiation Laboratory (NSRL) (https://cstr.cn/31131.02.HLS.XMCD.b) for the assistance of sXAS measurement.

**Additional information**

Supporting Information is available online or from the author.

**Correspondences:** Correspondences should be addressed: Prof. Xuanchi Zhou (*xuanchizhou@sxnu.edu.cn*), Prof. Guowei Zhou (*zhougw@sxnu.edu.cn*), and Prof. Xiaohong Xu (*xuxh@sxnu.edu.cn*).

**Figures and captions**

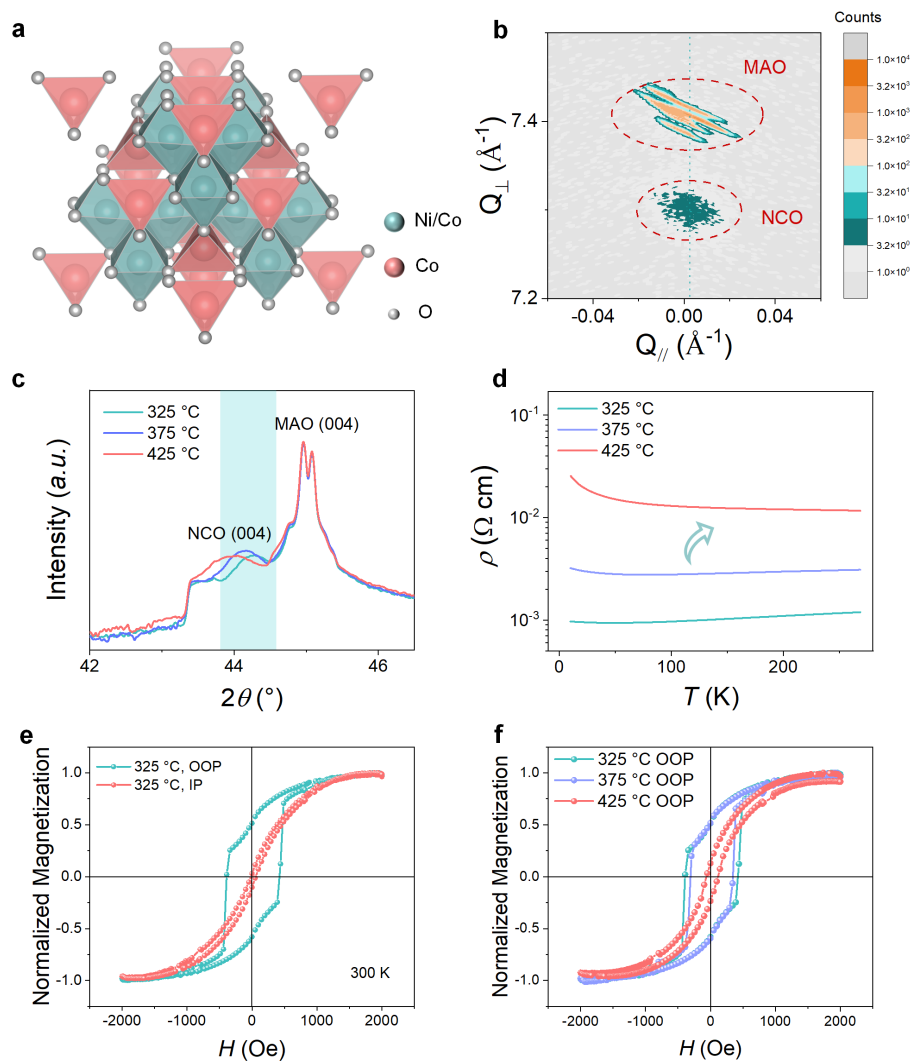

**Figure 1. Tailoring the ferrimagnetic metallic state in NiCo$_2$O$_4$ via cation disorder induced by growth temperature. a,** Schematic of the inverse spinel structure and inherent cation distribution disorder in NiCo$_2$O$_4$ system. **b,** Reciprocal space mapping (RSM) around the (226) reflection for NCO film grown at 325 °C. **c,** X-ray diffraction (XRD) patterns compared for the grown NCO/MAO heterostructures at different substrate temperature ranging from 325 °C to 425 °C. **d,** Temperature dependence of material resistivity ($\rho$-$T$) as measured for NCO films grown at different substrate temperature. **e.** The magnetization hysteresis (*M-H*) loops compared for NCO films deposited at 325 °C along the OOP and IP directions at 300 K. **f.** *Out-of-plane M-H* loops for NCO films grown at different temperatures.

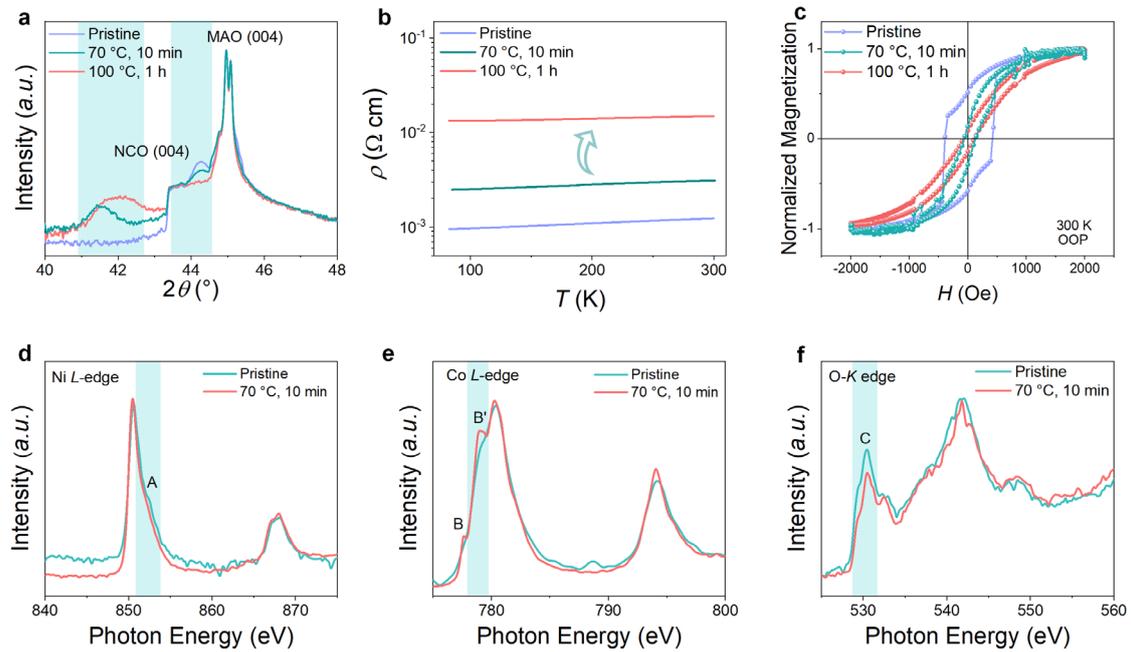

**Figure 2. Hydrogenation-induced structural and magnetoelectric state evolutions.
a,** XRD patterns compared for NCO films before and after hydrogenation. **b,** $\rho$-$T$ tendencies as measured for NCO/MAO heterostructures upon different hydrogenation treatments. **c,** Normalized *out-of-plane M-H* loops for NCO films through different hydrogenation treatment at 300 K. **d-f,** Soft X-ray absorption spectroscopy (sXAS) for the **d,** Ni *L*-edge, **e,** Co *L*-edge and **f,** O *K*-edge compared for NCO films through hydrogenation.

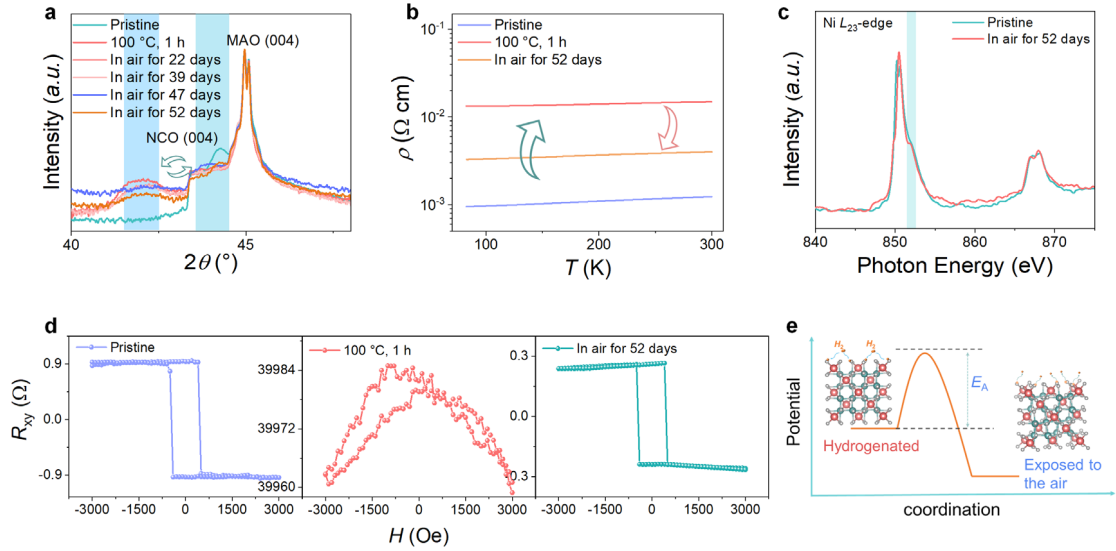

**Figure 3. Chemical stability and reversibility in hydrogenated electromagnetic states. a,** XRD patterns compared for NCO/MAO heterostructure upon hydrogenation and dehydrogenation. **b,** $\rho$-$T$ curves compared for NCO/MAO heterostructure upon hydrogenation and dehydrogenation. **c,** Comparing the Ni $L$-edge of sXAS spectra for pristine and dehydrogenated NCO films. **d,** The anomalous hall effect (AHE) in NCO films through hydrogenation and dehydrogenation. **e,** Schematic illustration of the spontaneous dehydrogenation mechanism in NCO.

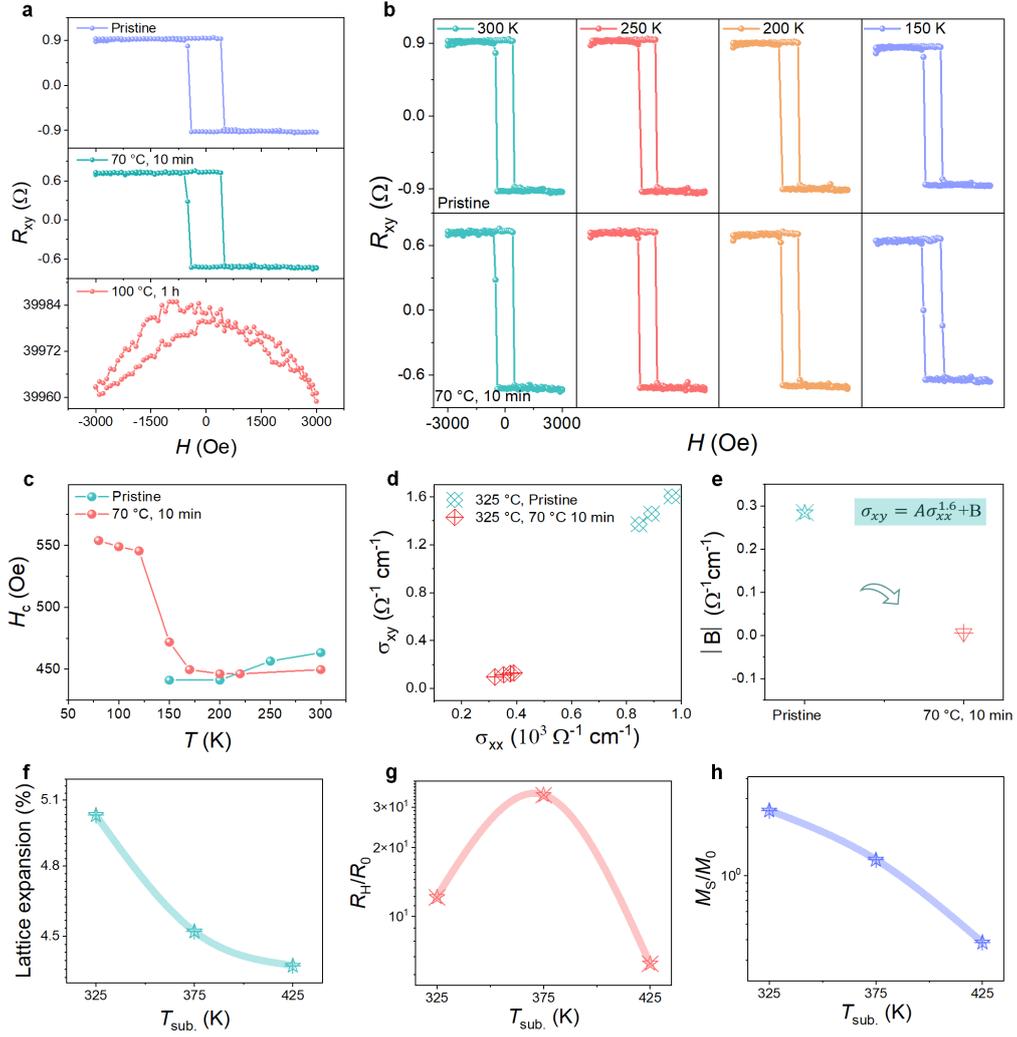

**Figure 4. Tunable electromagnetic transports in NCO through magneto-ionic control and cation disorder. a,** Evolution of the AHE in NCO films via hydrogenation. **b,** Comparison of the AHE in pristine and hydrogenated NCO films across different temperatures. **c,** Temperature-dependence of Coercive Field ($H_c$) for pristine and hydrogenated NCO films. **d,** Scaling relation between longitudinal and hall conductivity in NCO upon hydrogenation. **e,** Comparing the absolute magnitude of B in NCO associated with the intrinsic contribution from Berry curvature. **f-g,** Comparison of the regulation of **f,** structural evolution, **g,** electrical transport, and **h,** magnetic property in NCO films grown at different substrate temperature ($T_{sub.}$) through hydrogenation.